\pdfoutput=1
\documentclass[]{spie}  

\usepackage{amsmath,amsfonts,amssymb}
\usepackage{graphicx}
\usepackage[colorlinks=true, allcolors=blue]{hyperref}
\usepackage{siunitx}
\usepackage{booktabs}
\usepackage{color}
\usepackage{xcolor}
\usepackage{tablefootnote}

\newcommand{\bicep}{{\textsc {Bicep}}}
\newcommand{\bicepthree}{{\textsc {Bicep3}}}
\newcommand{\planck}{{\textit {Planck}}}
\newcommand{\wmap}{{WMAP}}

\newcommand{\keck}{{\textit{Keck Array}}}
\newcommand{\biceparray}{{\textsc{Bicep} Array}}

\newcommand{\ukcmbrts}{$\mu\mathrm{K}_{\mathrm{\mbox{\tiny\textsc cmb}}}\sqrt{\mathrm{s}}$}
\newcommand{\ukcmb}{$\mu\mathrm{K}_{\mathrm{\mbox{\tiny\textsc cmb}}}$}

\title{Receiver development for BICEP Array, a next-generation CMB polarimeter at the South Pole}

\author[a]{L.~Moncelsi}
\affil[a]{Department of Physics, California Institute of Technology, Pasadena, California 91125, USA}
\author[b]{P.~A.~R.~Ade}
\affil[b]{School of Physics and Astronomy, Cardiff University, Cardiff, CF24 3AA, United Kingdom}
\author[c,d]{Z.~Ahmed}
\affil[c]{SLAC National Accelerator Laboratory, 2575 Sand Hill Road, Menlo Park, CA 94025}
\affil[d]{Kavli Institute for Particle Astrophysics and Cosmology, Stanford University, 452 Lomita Mall, Stanford, CA 94305}
\author[e]{M.~Amiri}
\affil[e]{Department of Physics and Astronomy, University of British Columbia, Vancouver, British Columbia, V6T 1Z1, Canada}
\author[f]{D.~Barkats}
\affil[f]{Center for Astrophysics, Harvard \& Smithsonian, Cambridge, MA 02138, U.S.A}
\author[a]{R.~Basu Thakur}
\author[g]{C.~A.~Bischoff}
\affil[g]{Department of Physics, University of Cincinnati, Cincinnati, Ohio 45221, USA}
\author[a,h]{J.~J.~Bock}
\affil[h]{Jet Propulsion Laboratory, Pasadena, California 91109, USA}
\author[i,j]{V.~Buza}
\affil[i]{Kavli Institute for Cosmological Physics, University of Chicago, Chicago, IL 60637, USA}
\affil[j]{Department of Physics, Harvard University, Cambridge, MA 02138, USA}
\author[k]{J.~Cheshire}
\affil[k]{Minnesota Institute for Astrophysics, University of Minnesota, Minneapolis, 55455, USA}
\author[l,f]{J.~Connors}
\affil[l]{National Institute of Standards and Technology, Boulder, Colorado 80305, USA}
\author[f]{J.~Cornelison}
\author[m]{M.~Crumrine}
\affil[m]{School of Physics and Astronomy, University of Minnesota, Minneapolis, 55455, USA}
\author[d,c,n]{A.~Cukierman}
\affil[n]{Department of Physics, Stanford University, 382 Via Pueblo Mall, Stanford, CA 94305}
\author[l]{E.~V.~Denison}
\author[f]{M.~Dierickx}
\author[o]{L.~Duband}
\affil[o]{Service des Basses Temp\'{e}ratures, Commissariat \`{a} lEnergie Atomique, 38054 Grenoble, France}
\author[f]{M.~Eiben}
\author[e]{S.~Fatigoni}
\author[p,q]{J.~P.~Filippini}
\affil[p]{Department of Physics, University of Illinois at Urbana-Champaign, Urbana, Illinois 61801}
\affil[q]{Department of Astronomy, University of Illinois at Urbana-Champaign, Urbana, Illinois 61801, USA}
\author[d,n]{N.~Goeckner-Wald}
\author[f]{D.~C.~Goldfinger}
\author[n]{J.~Grayson}
\author[f]{P.~Grimes}
\author[k,n]{G.~Hall}
\author[e]{M.~Halpern}
\author[f]{S.~Harrison}
\author[c,d]{S.~Henderson}
\author[h,a]{S.~R.~Hildebrandt}
\author[l]{G.~C.~Hilton}
\author[l]{J.~Hubmayr}
\author[a]{H.~Hui}
\author[c,d,n,l]{K.~D.~Irwin}
\author[a,n]{J.~Kang}
\author[i,f]{K.~S.~Karkare}
\author[a]{S.~Kefeli}
\author[f,j]{J.~M.~Kovac}
\author[n,c,d]{C.~L.~Kuo}
\author[m]{K.~Lau}
\author[i]{E.~M.~Leitch}
\author[h]{K.~G.~Megerian}
\author[a]{L.~Minutolo}
\author[m,n]{Y.~Nakato}
\author[r,n]{T.~Namikawa}
\affil[r]{Department of Applied Mathematics and Theoretical Physics, University of Cambridge, Cambridge CB3 0WA, UK}
\author[h]{H.~T.~Nguyen}
\author[h,a]{R.~O'Brient}
\author[g]{S.~Palladino}
\author[m]{N.~Precup}
\author[o]{T.~Prouve}
\author[k,m]{C.~Pryke}
\author[f]{B.~Racine}
\author[l]{C.~D.~Reintsema}
\author[a]{A.~Schillaci}
\author[f]{B.~L.~Schmitt}
\author[a]{A.~Soliman}
\author[f,j]{T.~St.~Germaine}
\author[a]{B.~Steinbach}
\author[b]{R.~V.~Sudiwala}
\author[d,n]{K.~L.~Thompson}
\author[b]{C.~Tucker}
\author[h]{A.~D.~Turner}
\author[p,g]{C.~Umilt\`{a}}
\author[s,i]{A.~G.~Vieregg}
\affil[s]{Department of Physics, Enrico Fermi Institute, University of Chicago, Chicago, IL 60637}
\author[a]{A.~Wandui}
\author[h]{A.~C.~Weber}
\author[e]{D.~V.~Wiebe}
\author[m]{J.~Willmert}
\author[c,d,n]{W.~L.~K.~Wu}
\author[n]{E.~Yang}
\author[n,c,d]{K.~W.~Yoon}
\author[d,c,n]{E.~Young}
\author[n]{C.~Yu}
\author[f]{L.~Zeng}
\author[a]{C.~Zhang}
\author[a]{S.~Zhang}

\authorinfo{email: \url{moncelsi@caltech.edu}}

\begin{document} 
\maketitle

\begin{abstract}

A detection of curl-type ($B$-mode) polarization of the primary CMB would be direct evidence for the inflationary paradigm of the origin of the Universe. The \bicep/\keck\ (BK) program targets the degree angular scales, where the power from primordial $B$-mode polarization is expected to peak, with ever-increasing sensitivity and has published the most stringent constraints on inflation to date. \biceparray\ (BA) is the Stage-3 instrument of the BK program and will comprise four \bicep3-class receivers observing at 30/40, 95, 150 and 220/270\,GHz with a combined 32,000+ detectors; such wide frequency coverage is necessary for control of the Galactic foregrounds, which also produce degree-scale $B$-mode signal. The 30/40\,GHz receiver is designed to constrain the synchrotron foreground and has begun observing at the South Pole in early 2020. By the end of a 3-year observing campaign, the full \biceparray\ instrument is projected to reach $\sigma_r$ between 0.002 and 0.004, depending on foreground complexity and degree of removal of $B$-modes due to gravitational lensing (delensing). This paper presents an overview of the design, measured on-sky performance and calibration of the first BA receiver. 
We also give a preview of the added complexity in the time-domain multiplexed readout of the 7,776-detector 150\,GHz receiver.

\end{abstract}


\section{Introduction}

Measurements of the polarization of the Cosmic Microwave Background (CMB) provide crucial information to further our understanding of the early Universe. The Lambda cold dark matter ($\Lambda$CDM) model predicts an $E$-mode polarization pattern in the CMB at the level of a few $\mu$K as well as a $B$-mode polarization signal arising from gravitational lensing of $E$-modes by the large-scale structure of the Universe. Inflationary gravitational waves are predicted to produce an excess of $B$-mode power on degree angular scales that is scaled by the tensor-to-scalar ratio $r$. The value of $r$ is directly tied to the energy scale of inflation~\cite{kamionkowski2016quest}. While certain classes of inflation models predict extremely small values of $r$, a detection of primordial $B$-mode polarization would be direct evidence for the theory of inflation. However, polarized Galactic dust and synchrotron foregrounds hamper our ability to measure primordial gravitational waves and need to be disentangled from the degree-scale $B$-mode polarization signal by observing the sky at multiple frequencies with exquisite sensitivity. Additionally, the removal of $B$-mode power due to gravitational lensing (delensing) can lead to improved constraints on $r$, but it requires higher resolution observations~\cite{Smith2012}.

\begin{figure}[h]
\centering
\includegraphics[width=0.75\textwidth]{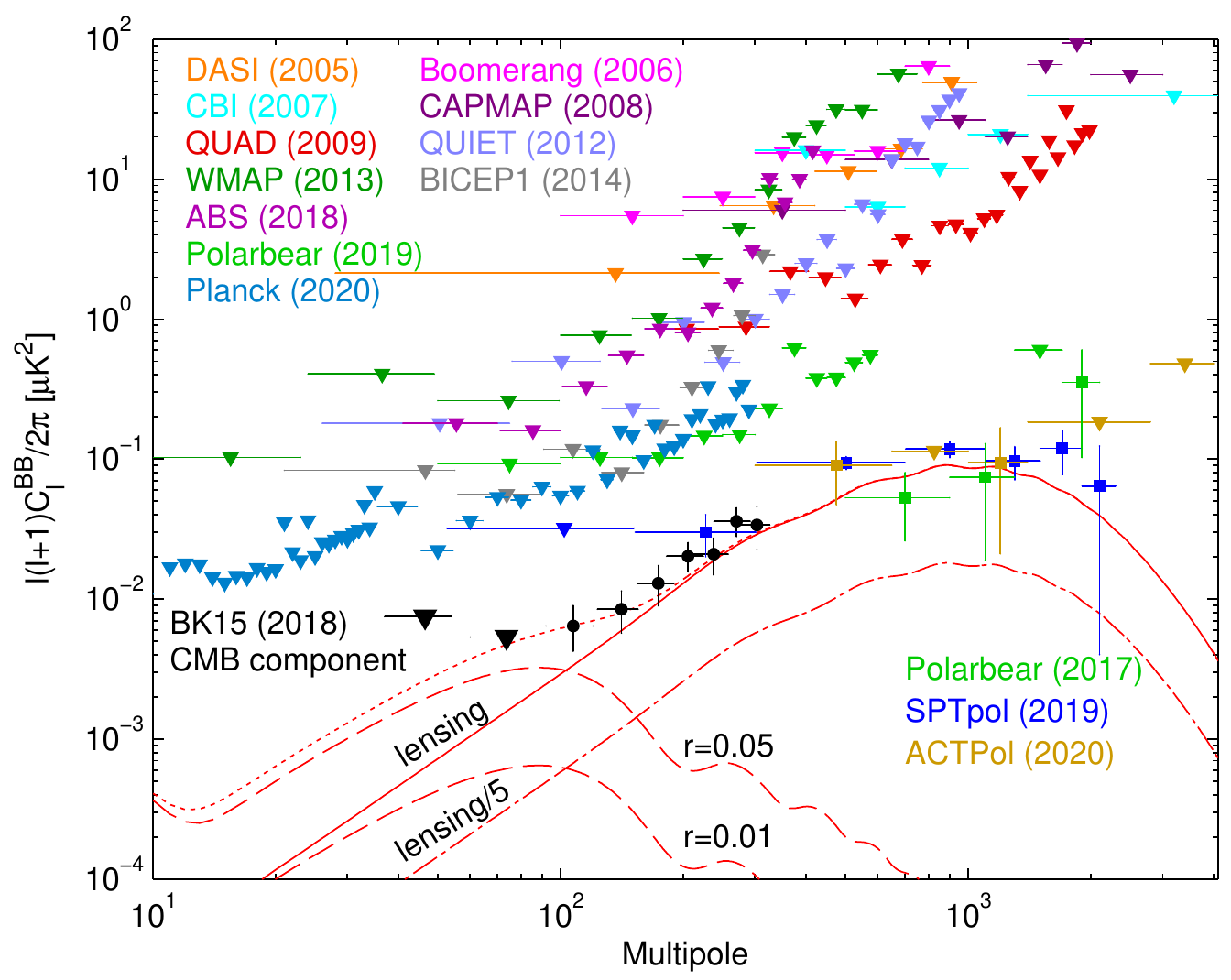}
\caption[Published $B$-mode polarization measurements]{$B$-mode polarization measurement by different experiments as of Nov. 2020. The \planck\ points are on arXiv~\cite{Planck_r_2020}.
}
\label{fig:bk_15_status}
\index{figure}
\end{figure}

\begin{figure}
   \begin{center}
   \includegraphics[width=1.0\textwidth]{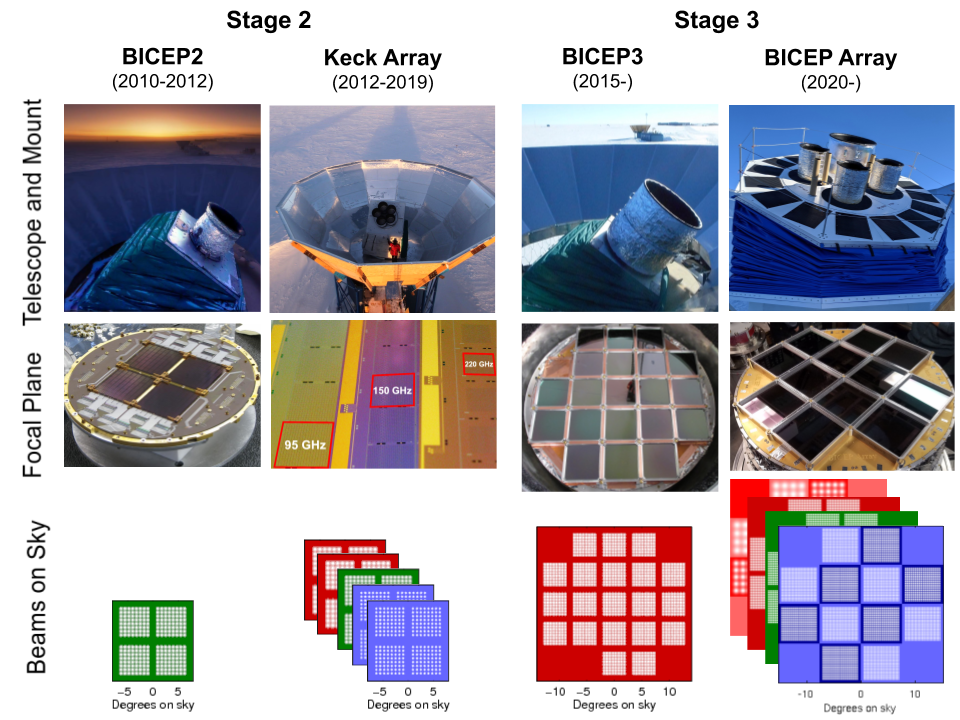}
   \end{center}
   \caption{The progression of the \bicep/\keck\ program leading to the \biceparray.
   Bottom row: the beam patterns of the focal planes on the sky shown on a common scale.
   Each square represents a single receiver, and the colors indicate different observing frequencies: light red for 30/40\,GHz, red for 95\,GHz, green for 150\,GHz, and blue for 220/270\,GHz.
   }
   \label{fig:bk_program} 
   \end{figure}

The most stringent published constraint on the tensor-to-scalar ratio is $r_{0.05}<0.06$ at 95\% confidence from \bicep/\keck\ data in conjunction with \planck\ and \wmap\ measurements~\cite{BK15} (see Figure~\ref{fig:bk_15_status}, which includes the most recent \planck\ points~\cite{Planck_r_2020}). Over the past 15 years, our experimental strategy of designing small-aperture, cryogenic, refracting wide-field telescopes observing from the South Pole has proven successful in probing the degree-scale polarization of the CMB (multipoles 35 $<$ \emph{l} $<$ 300) with tight control of systematic errors. \bicep1\ was the first polarimeter designed specifically to target the $B$-mode signal and operated from January 2006 through December 2008 with 49 orthogonal pairs of polarization-sensitive NTD bolometers observing at 100 and 150\,GHz. \bicep2\ observed the sky with 500 antenna-coupled transition-edge sensor (TES) bolometers at 150\,GHz from 2010 to 2012, and reported an excess of $B$-mode power over the base lensed-$\Lambda$CDM model in the range $30<l<150$~\cite{ade2014detection}, which was later attributed to polarized dust emission~\cite{BicepKECKxPlanck}. The \keck\ consisted of five \bicep2\,-like 260\,mm aperture receivers and started observations at 150\,GHz in 2012. 
The interchangeable \keck\ receivers allowed us to diversify the frequency coverage over the years: 95\,GHz receivers were deployed in 2014, 220\,GHz in 2015, and 270\,GHz in 2017. The higher-throughput \bicepthree\ receiver replaced \bicep2\ in its mount in 2015, with a 520\,mm aperture and 2500 detectors all operating at 95\,GHz for an on-sky instantaneous instrument sensitivity of 6.7\,\ukcmbrts \cite{Kang18}. Figure~\ref{fig:bk_program} shows the progression of the \bicep/\keck\ program, to larger apertures, larger focal planes, and wider frequency coverage.

\begin{figure}
  \centering
  \includegraphics[width=1\textwidth]{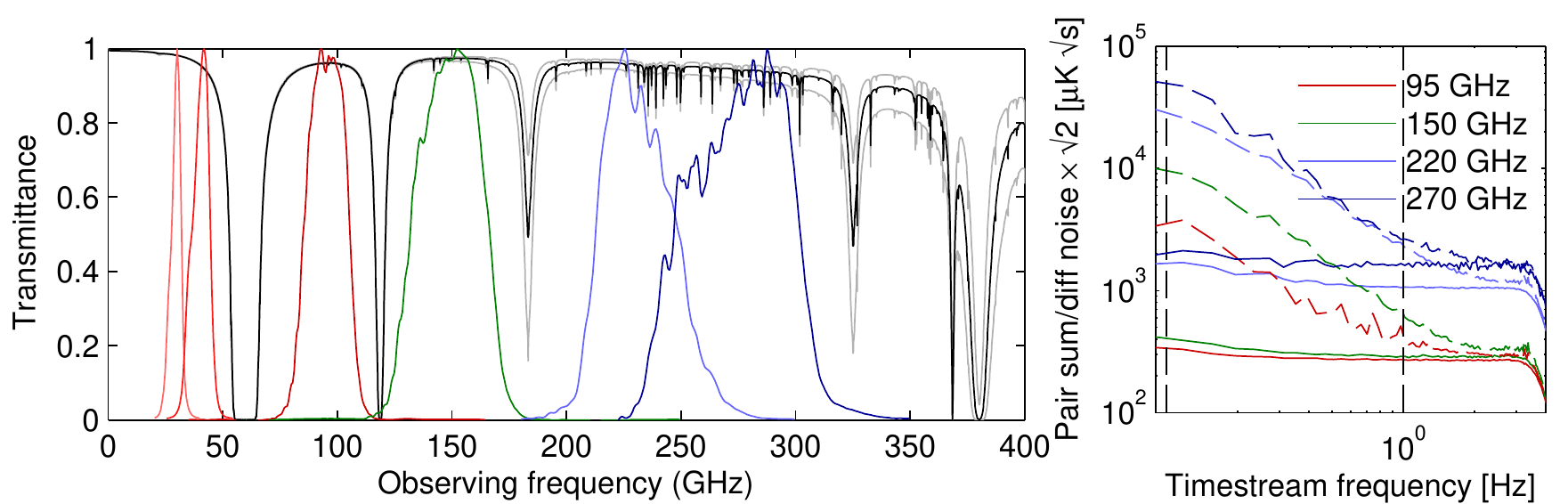}
  \caption{
  {\it Left:} Comparison of atmospheric transmission at the South Pole  with the bandpasses of \bicep/\keck\ and \biceparray.
  Median atmospheric transmission during the observing season is shown in black, bracketed by the 10\textsuperscript{th} and 90\textsuperscript{th} percentiles. Transmission drops only slightly across 200--300\,GHz, making dust observations in the upper part of this window effective, with similar dust sensitivity to the 220 GHz band. {\it Right:} Minimally processed timestream pair-sum and pair-difference noise spectra from \keck.
The stable Antarctic atmosphere enables observations at all of these frequencies that are low-noise across the indicated science band from 0.1--1\,Hz, corresponding to $25\lesssim\ell\lesssim250$.
}
  \label{fig:atmos}
  \index{figure}
\end{figure}

\begin{table}
\small
\begin{center}
\begin{tabular}[c]{|l|c|c|c|c|}
\hline
Receiver       & Nominal  & Nominal Single       & Beam       & Survey Weight \\
Observing Band & Number of& Detector NET         & FWHM       & Per Year      \\
(GHz)          & Detectors& (\ukcmbrts)          & (arcmin)   & (\ukcmb)$^{-2}$ yr$^{-1}$\\
\hline\hline   						
\keck    &          &                      &            &               \\
\ \ \ 95           & \textbf{288} & \textbf{288}                  & \textbf{43}    & \textbf{24,000}   \\
\ \ \ 150          & \textbf{512} & \textbf{313}                  & \textbf{30}    & \textbf{30,000}   \\
\ \ \ 220          & \textbf{512} & \textbf{746}                  & \textbf{21}    & \textbf{2,750}    \\
\ \ \ 270          & \textbf{512}      & 1,310                 & \textbf{17}         & 800           \\
\hline\hline   						
\bicepthree    &          &                      &            &               \\
\ \ \ 95           & \textbf{2,560}& \textbf{265}                  & \textbf{24}    & \textbf{240,000}       \\
\hline\hline   						
\biceparray     &          &                      &            &               \\
$\big \langle \hspace{-3pt} \begin{array}{l} 30 \\ 40 \end{array}$   & $\begin{array}{l} 192 \\ 300 \end{array}$
  & $\begin{array}{l} 260 \\ 318 \end{array}$ & $\begin{array}{l} 76 \\ 57 \end{array}$
  & $\begin{array}{l} 19,500 \\ 20,500 \end{array}$       \\
\ \ \ $95$           & $4,056$     & $265$                  & $24$         & $380,000$       \\
\ \ \ $150$          & $7,776$     & $313$                  & $15$         & $455,600$       \\
$\big \langle \hspace{-3pt} \begin{array}{c} 220 \\ 270 \end{array}$   & $\begin{array}{c} 8,112 \\ 12,288 \end{array}$
  & $\begin{array}{c} 746 \\ 1,310 \end{array}$ & $\begin{array}{c} 11 \\ 9 \end{array}$
  & $\begin{array}{c} 58,600 \\ 19,200 \end{array}$       \\
\hline\hline
\end{tabular}
\end{center}
\caption[Receiver parameters and sensitivity for \bicep\ program]{Receiver parameters as used in sensitivity projections. Boldface numbers are actual/achieved quantities for existing receivers. The remaining values in the survey weight column are scaled from the achieved survey weights using only the ratio of the number of detectors, plus, if necessary to change frequency, the ratio of nominal NET values squared.}
\label{tab:rxs}
\end{table}


\section{The Bicep Array Instrument}

The latest instrument in our program is \biceparray~\cite{Hui18,Schillaci19}, which adopts the same interchangeable concept used in \keck\ and is comprised of four \bicepthree-class receivers in six frequency bands spanning from 30 to 270\,GHz (Figure~\ref{fig:atmos}), which are all 27\% wide. The lowest (30/40\,GHz) and highest (220/270\,GHz) frequency receivers are dual-band receivers each featuring a checker-board of focal plane modules with single-band detectors centered at the two adjacent bands (Figure~\ref{fig:bk_program}). Splitting the atmospheric windows this way provides more spectral information on the Galactic synchrotron and dust emission, thus improving the constraints on the foreground model parameters. Scaling from previous on-sky performance, Table~\ref{tab:rxs} shows the expected sensitivities and survey weights\footnote{Survey Weight is a single number representing the total raw experimental sensitivity achieved, and is defined as $W = 2A/N^2$ where $A$ is the map area and $N$ is the noise level in the $Q/U$ maps. Survey weight is useful because it scales linearly with integration time, number of detectors, and statistical sensitivity to $r$~\cite{Buza2020}.} of the 4 receivers, which amount to 32,000+ photon-noise limited polarization-sensitive detectors.

At the end of 2019, the \keck\ telescope mount has been replaced by the new \biceparray\ mount~\cite{Crumrine18}, which provides bore-sight rotation about the optical axis, in addition to rotation in Azimuth and Elevation, following the \keck\ and \bicepthree\ designs. During the same Austral Summer, we also deployed the first 30/40\,GHz \biceparray\ receiver, equipped with a full complement of detector modules to fill the focal plane, which has now observed for an entire season. The other \biceparray\ receivers will be installed in the new mount with a staged approach over the next few years\footnote{Due to the COVID-19 pandemic, no crew will deploy at the end of 2020 and the instrument will remain unchanged.}. In the meantime, we are continuing observations with a subset of the \keck\ receivers installed in open slots until they are filled by available \biceparray\ receivers. The observations will continue to focus on the deep \bicep/\keck\ sky patch so that the new low-frequency data can be combined with existing multi-frequency observations. An updated project timeline is shown in Figure~\ref{fig:bk_projections_2018}.

%

\subsection{Receiver Overview}

\begin{figure}
  \centering
\includegraphics[width=1.0\textwidth]{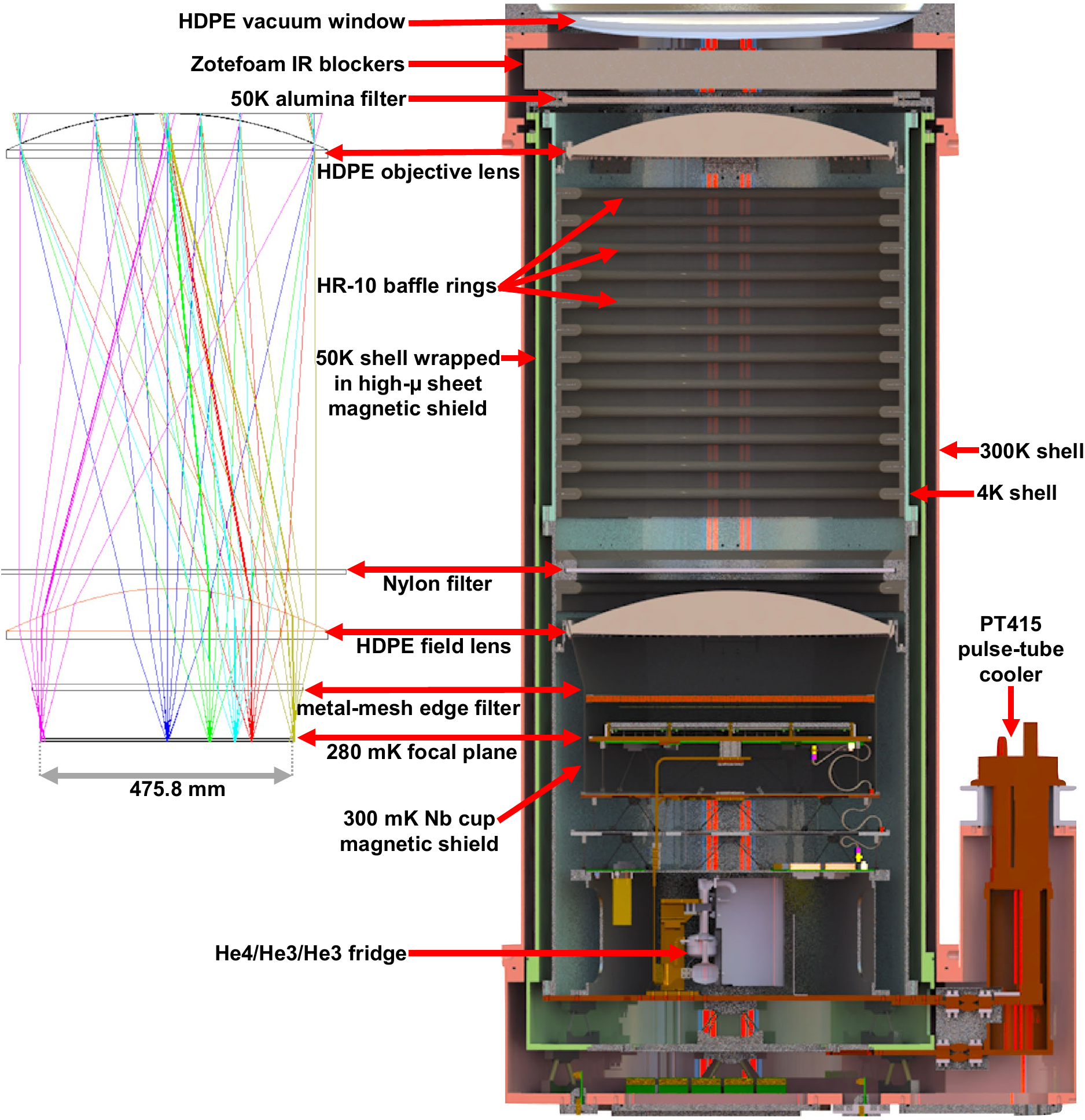}
  \caption[\biceparray\ optical diagram]{
\biceparray\ optical diagram and cross-section of the cryogenic receiver. All optical components, except for the Zotefoam filters, are anti-reflection coated to minimize in-band reflections. The HDPE lenses, nylon filter, and aperture stop are precision-mounted in the receiver and cooled to 4\,K to reduce detector loading. The focal plane assembly, which houses 12 detector modules, is cooled to 280\,mK by a sorption fridge and surrounded by a superconducting Nb magnetic shield. The focal plane is actively temperature-controlled to achieve detector baseline stability over timescales of hours. The radially-symmetric receiver design allows well-matched beams for orthogonally polarized detectors at the focal plane.
}
  \label{fig:ba_raytrace}
  \index{figure}
\end{figure}

\biceparray\ is largely based on the successful design of \bicepthree \cite{ahmed2014bicep3,grayson2016bicep3,hui2016bicep3,wu2016initial}. Each receiver is housed in a custom-designed vacuum cryostat~\cite{Crumrine18} 2.1\,m tall and 0.9\,m in diameter (excluding the additional envelopes of the multi-channel readout electronics and the pulse-tube cooler). Figure~\ref{fig:ba_raytrace} shows a cross-section of the receiver, along with the optical diagram for reference. The vacuum and 50\,K shells are designed with a short base-stage, and long cylinders for the main section, which allow users to lift off the outer shells from the base plate without undoing the cabling, thermal joints and sub-Kelvin truss structure. The vacuum window is made out of HDPE for the 30/40\,GHz receiver, while thinner solutions for the higher frequencies are actively being pursued~\cite{barkats2018window}.

 
The 4\,K shell is divided into two lengthwise segments for ease of access during integration; the top optics section houses the objective lens and baffle rings designed to minimize reflections within the tube. The lower optics section holds the nylon filter and the field lens. The 4\,K plate supports the fridge space and the sub-kelvin thermo-mechanical truss structure. The 50\,K volume is supported by G-10 truss legs, providing robust structural support while maintaining low thermal conductivity between temperature stages. The truss legs for the 4\,K and sub-Kelvin structures are instead made out of carbon fiber, owing to its superior room-temperature elastic modulus to low-temperature thermal conductivity ratio at $\leq 4$\,K\cite{Runyan2008}.

\subsection{Thermal Architecture}

All \biceparray\ receivers (see Figure~\ref{fig:ba_raytrace}) use a Cryomech PT415\footnote{\url{https://www.cryomech.com/products/pt415/}} pulse tube with nominal cooling capacity of 40\,W and 1.5\,W at the 50\,K and 4\,K stages, respectively. The pulse tube is driven by a remote stepper motor that is physically detached and electrically/mechanically isolated from the cold head. Residual mechanical vibrations induced by the fittings connecting the head and the motor are damped by a spring bellows connecting the cold head to the cryostat shell. We also use e-beam welded heat straps with flexible sections of ropelay stranded copper in order to minimize the rigidity and the transmission of vibration to the cryostat stages.

The optical filtering follows the successful \bicepthree\ scheme \cite{ahmed2014bicep3}. Behind the 560\,mm-aperture vacuum window, which lets $>100$\,W of thermal radiation into the receiver, a stack of 12 1/8"-thick layers of Zotefoam HD-30\footnote{\url{https://www.zotefoams.com/wp-content/uploads/2016/02/HD30-December-2017.pdf}} with 1/10" spacers provides infrared filtering~\cite{Choi2013,2017arXiv170602464A}, in conjunction with a 10\,mm-thick IR-absorptive alumina filter\cite{inoue2014cryogenic}, which is installed at the top of the 50\,K stage. The lower 4\,K optics section holds an absorptive 7.45\,mm-thick nylon filter, to prevent any short-wavelength radiation let through by the alumina filter above from reaching the focal plane. Finally a low-pass metal-mesh filter~\cite{ade2006review,tucker2006thermal} installed at 300\,mK defines the upper edge of the band.

Combining measurements with a thermal model, the total loading on the 50\,K and 4\,K stages with the receiver on-sky is estimated to be about $\sim$40\,W and $\sim$0.5\,W, respectively. Table~\ref{tab:2020_temps} reports the median \biceparray\ system temperatures during CMB observations in 2020.



Sub-Kelvin cooling for the detectors is provided by a three-stage helium (4He/3He/3He) sorption fridge from CEA Grenoble\cite{ICC_Duband}, with heat intercepts at 2\,K (4He stage), 300\,mK (intermediate cooler, or IC), and 250\,mK (ultra cooler, or UC). With a design total loading of 70 and 15\,$\mu$W at the 300 and 250\,mK stages, respectively, the fridge has sufficient cooling capacity to maintain a 3-day uninterrupted observing schedule between thermal cycles. With the telescope on-sky during observations, we have achieved a minimum fridge hold time of 2.5\,days, after which other scheduling constraints become relevant.

The sub-Kelvin structure is located on top of the 4\,K base plate, above the sorption fridge. It is separated into three thermal, ``wedding-cake'' carbon-fiber truss stages at 2\,K, 300\,mK, and 280\,mK. Each stage provides radiative shielding and room for cable heat-sinking to the respective fridge stages, allowing low-loading environment for the focal plane and detectors. The detector modules and copper focal plane plate are heat-sunk to the fridge UC stage via a flexible high-purity copper-foil heat strap. The strap connects to the focal plane through a stainless steel block. The temperature of the strap is actively regulated, and the block serves as a passive low-pass thermal filter to attenuate thermal noise from the control circuit. The estimated conductive loading on the 2\,K, IC, and UC stages is $\sim$178, $\sim$55, and 0.234\,$\mu$W, respectively, from heat-sinking of cables and the mechanical truss structure, but not including radiative contributions (see Table~\ref{tab:ba_subK_load}). 

\begin{table}
  \center
  \begin{tabular}{l c c c} 
    \multicolumn{4}{c}{Truss structure conducted load budget}\\
    \toprule 
    					&	2\,K ($\mu$W)	&	300\,mK ($\mu$W)	&	250\,mK (nW)\\
    \toprule 
    18 100-way NbTi cables (TDM)		&	84.1		&	18.8			&	82.2	\\
    4 25-way Manganin cables	(HK) &	1.9			&	0.5		&	2.8		\\
    Cernox cables (HK)		&	6.6		&	1.2				&	2.4		\\
    Heater cables (HK)		&	2.8			&	0.5				&	1.8		\\
    Carbon fiber trusses	(M)	&	33.6			&	12.9			&	47.1	\\
    Aluminized mylar RF shield (M)	&	48.7			&	21.5			&	97.8	\\
    \midrule
    Total				&	177.7		&	55.5				&	234.1	\\
    \bottomrule
    \vspace{1mm}
  \end{tabular}
  \caption[Sub-Kelvin loading from conduction for \biceparray.]{Sub-Kelvin loading from conduction for \biceparray. This calculation~\cite{Olson1993,Woodcraft2010} is based on the maximum detector readout NbTi cable count for the highest-frequency TDM receiver (150\,GHz, see Table~\ref{tab:ba_tmux}). The housekeeping (HK) and miscellaneous (M) entries are common to all \biceparray\ receivers. More details on RF shielding are given in Section~\ref{rf_shield}.}
  \label{tab:ba_subK_load}
\end{table}

\subsection{Optics}
Following the same concept as in previous small-aperture telescopes in the \bicep/\keck\ series, the \biceparray\ receivers are simple diffraction-limited two-lens refractors, which provide a telecentric and flat focal surface with minimum aberrations over a wide field of view (see Figure~\ref{fig:ba_raytrace} for the optical diagram). The telescope has a mean $f$-ratio of $f$/1.57 at the center of the focal plane, which has a diameter of 475.8\,mm. The HDPE lenses are 650\,mm in diameter, with a clear aperture of 550\,mm and a field of view of 29.6\,degrees. The lenses and nylon filter are cooled to 4\,K to minimize loading on the detectors and are anti-reflection coated to minimize in-band reflections. The 4\,K space between the objective and the field lens is lined with baffle rings coated with Eccosorb\textsuperscript{\textregistered} HR-10 microwave absorber to suppress far-sidelobes reflections. The same Eccosorb\textsuperscript{\textregistered} is also used to define the 550\,mm optical stop for the system, which is located just above (sky side) the objective lens. The detector beams are designed to terminate on the cold stop with an edge taper ranging from -10 to -13\,dB with respect to the main lobe, depending on the frequency. Following the approach of previous optical designs in the \bicep/\keck\ telescope series, we take advantage of the excellent directivity provided by the uniform illumination of the planar phased-array antennas and maintain a conservative edge taper at the aperture stop~\cite{ade2015antenna}.

The current material for the vacuum window is 1"-thick HDPE at 30/40\,GHz, where HDPE has negligible in-band emission. At higher frequencies, thinner windows are being pursued, which will reduce the in-band optical loading and thereby significantly improve mapping speed\cite{barkats2018window}. 


\subsection{Detector Modules and Readout}
\label{sec:detectors}

\begin{figure}
  \centering
  {\includegraphics[width=.53\columnwidth, clip=true]{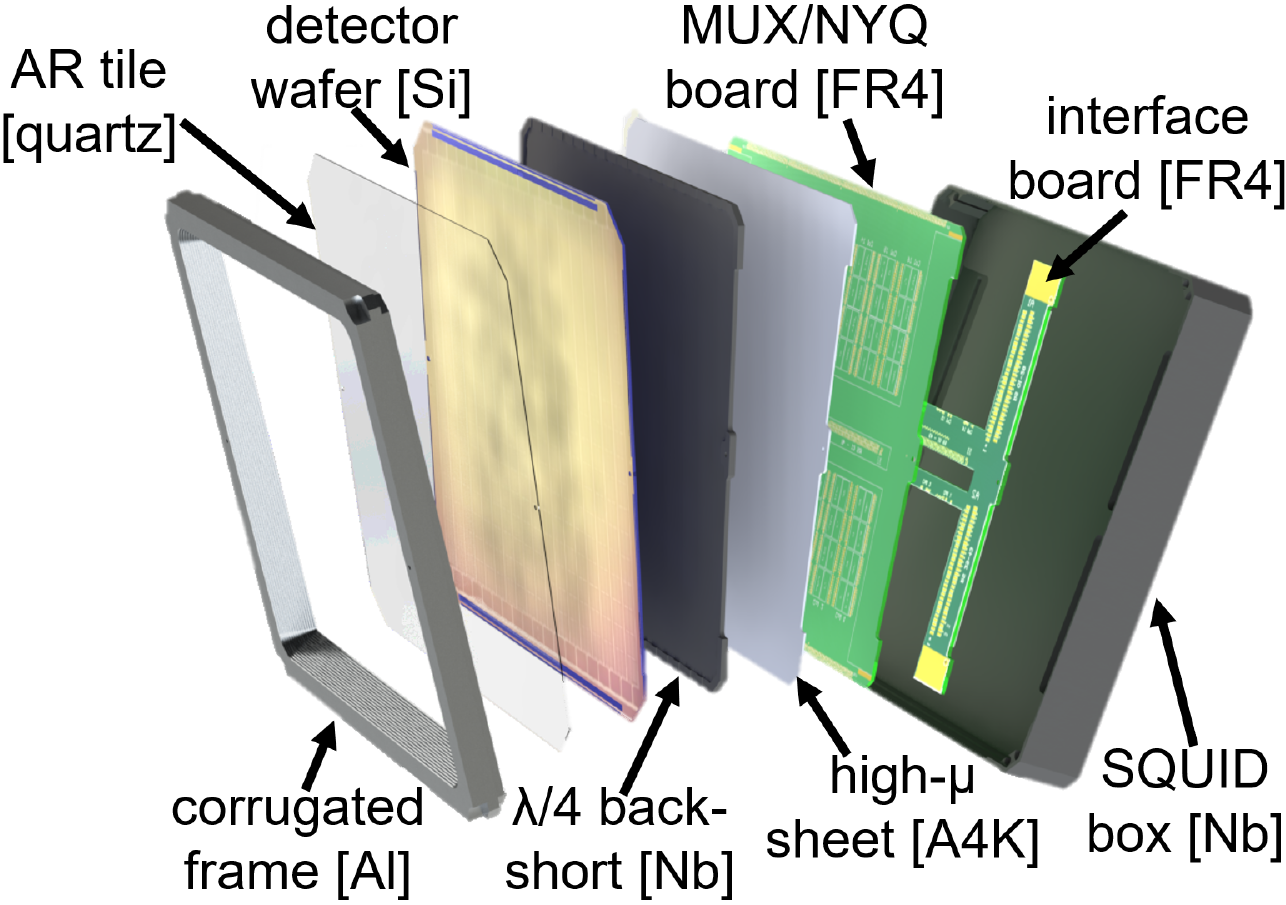}}
  {\includegraphics[width=.4635\columnwidth, clip=true]{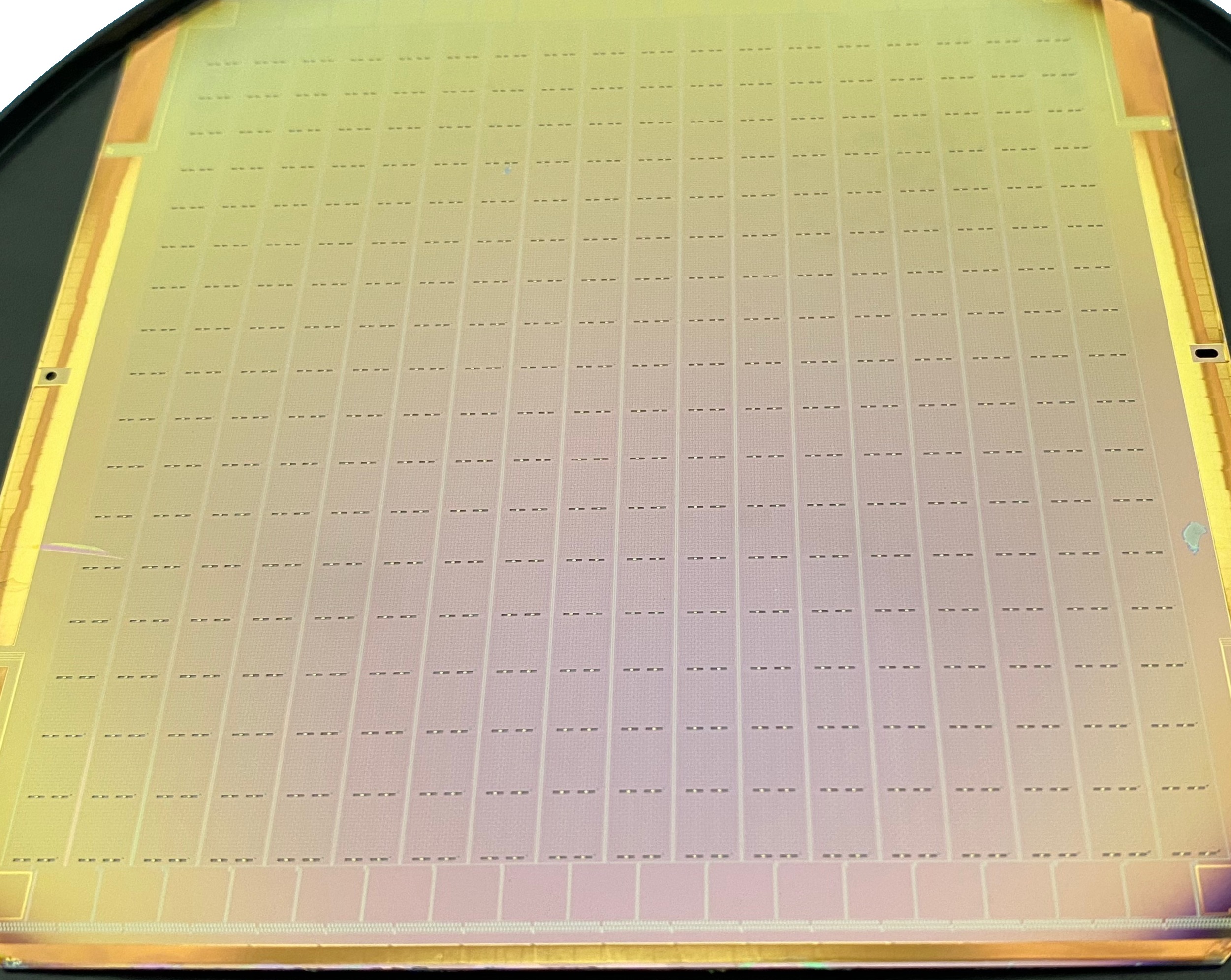}}
  \caption{{\it Left}: An exploded view of the \biceparray\ 150\,GHz module. The design is based on \bicepthree\ \cite{hui2016bicep3} but scaled to house a larger detector wafer. The SQUIDs are mounted on the ``MUX/NYQ'' board and completely enclosed in a superconducting Nb box for magnetic shielding. The ``interface'' board allows for four flexible copper cables to route the TDM row and column signals out of thin slots in the Nb box and into the focal plane distribution board (see Figure~\ref{fig:ba_readout_board}). The ``MUX/NYQ'' board for the BA 95 and 150\,GHz receivers is made out of FR4 with superconducting tin-copper traces, while \bicep3 and \biceparray\ 30/40\,GHz used aluminum traces on alumina substrates instead. {\it Right}: photo of a \biceparray\ 150\,GHz wafer with an 18x18 grid of phased-array slot antennas, each with dual-polarization detectors (648 total).}
  \label{fig:150GHz_module}
  \index{figure}
\end{figure}

\begin{figure}
  \centering
  \includegraphics[width=1\textwidth]{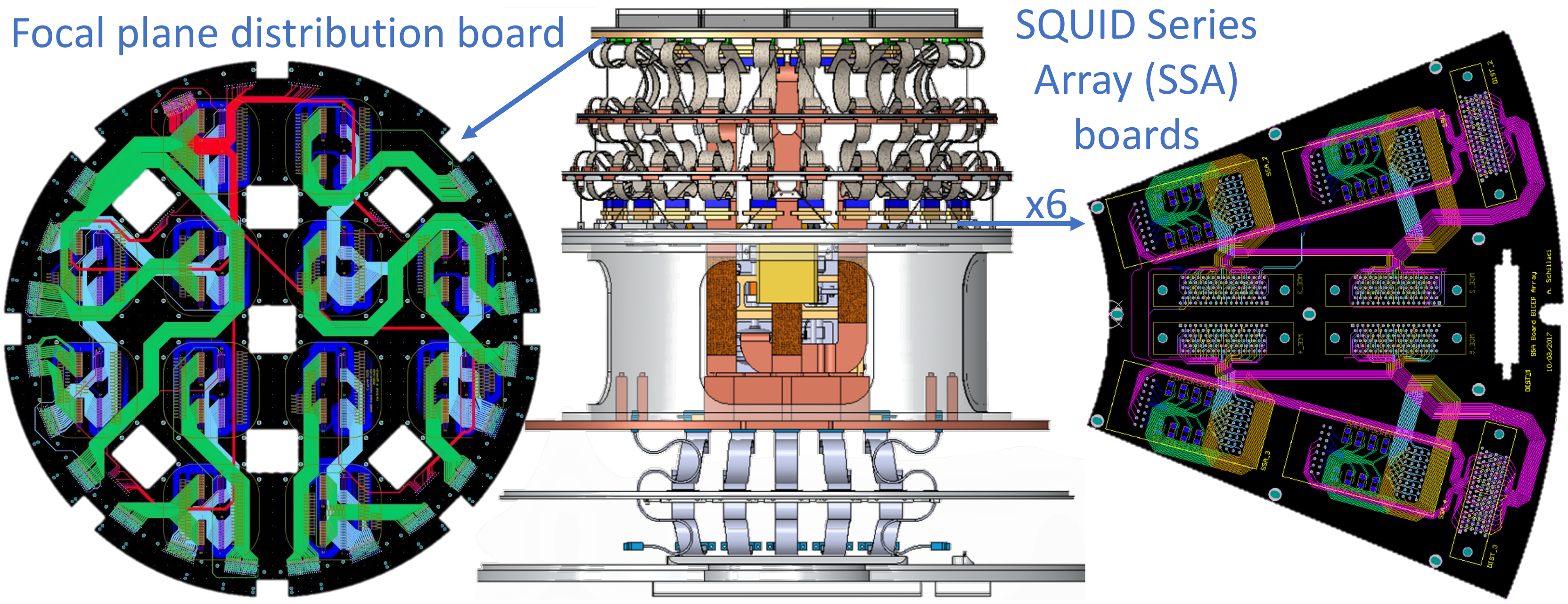}
  \caption{The focal plane distribution board gathers all the readout cabling from warmer stages and distributes them to the corresponding detector module at 280\,mK. The SSA board is located at 4\,K, and houses the SQUID Series Arrays (SSA). Each MCE connects to a single SSA board via five 100-way cables. The 30/40, 95, and 150\,GHz receivers have one, three and six SSA boards, respectively. Shown above is the 150\,GHz configuration.
}
  \label{fig:ba_readout_board}
  \index{figure}
\end{figure}

The \biceparray\ receivers are equipped with state-of-the-art dual-polarization antenna-coupled TES arrays, fabricated on 150\,mm diameter silicon wafers~\cite{Bonetti2009,ade2015antenna}. The detector arrays are based on planar in-phase combined slot antennas that provide higher pixel packing density than feedhorns, given a chosen edge taper~\footnote{See \url{https://cmb-s4.org/wiki/images/2019_10_16_CMBS4_DetectorTechUpdate_V01.pdf\#page=6}.}. The readout architecture is based on Time-Domain Multiplexing (TDM~\cite{Battistelli2008a,Battistelli2008b}), developed by the University of British Columbia and NIST, with two stages of SQUIDs driven using a Multi-Channel Electronics (MCE) system that provides SQUID/TES bias voltages, controls SQUID addressing, and digitizes detector signals. The readout chain is based on numerous previous successful experiments~\cite{Bintley2010,Henderson2016}, with the detector modules at 280\,mK carrying the first-stage SQUIDs on a PCB located behind the $\lambda$/4 backshort (see Figure~\ref{fig:150GHz_module}), which then connect to a custom focal plane distribution board and through superconducting NbTi wires to a stage of Series Array SQUIDs at 4\,K (see Figure~\ref{fig:ba_readout_board}). In the design of these PCBs and the architecture of the cables, we paid particular attention to pairing signal and return in order to minimize cross-talk among channels. Moreover, the SQUIDs and TES detectors are carefully shielded from sources of magnetic fields, including the Earth's field (see Section~\ref{magnetic_shield} for more details). 

Twelve detector modules are tiled onto the focal plane, each containing 32 to 2048 detectors, depending on the observing frequency (see Table~\ref{tab:rxs}). The module is based on the successful \bicepthree\ design\cite{hui2016bicep3}, except the detectors are now fabricated on 6" silicon wafers, instead of the previous 4" wafers to expedite lithographic fabrication with higher detector count per module. 
The first-stage SQUID readout multiplexing chips are housed in a FR4 circuit board inside the module with superconducting tin-copper traces (\bicep3 and \biceparray\ 30/40\,GHz used aluminum traces on alumina substrates instead). The housing is constructed with superconducting niobium and aluminum, which, along with a high-$\mu$ A4K sheet inside the module, are designed to achieve high magnetic shielding performance (Section~\ref{magnetic_shield}). Figure~\ref{fig:150GHz_module} shows the exploded view of the module design at 150\,GHz.

%
%

Electromagnetic interactions between the edge antennas and the metal frame around the module can cause polarized beam mismatch on the edge pixels, resulting in potential temperature-to-polarization leakage in the CMB maps. We designed corrugated walls to minimize this effect. The 30/40\,GHz module implements a novel broadband corrugation design that mitigates the differential ellipticity caused by the metal frame over 57\% bandwidth, from 25\,GHz to 45\,GHz~\cite{Soliman18,Soliman19}.

The first 30/40\,GHz receiver (BA1) in its final configuration will be populated with 6 modules at each frequency in a checker-board arrangement. The design for the two frequencies is similar, with appropriate dimensions of the antenna scaled accordingly to frequency~\cite{Zhang19}. Due to the COVID-19 pandemic, no crew will deploy at the end of 2020 and the instrument will observe the sky for two full seasons in the configuration it was first deployed at the end of 2019, with the following breakdown of modules: 4 at 30\,GHz, 7 at 40\,GHz, plus one prototype dual-band 30/40\,GHz module with 16 dichroic optical pixels (64 detectors; Shiu et al. in prep.~\footnote{\url{https://microdevices.jpl.nasa.gov/interactivepdfs/index.html?pdf=2020-MDL-Annual-report.pdf}, page 61.}). 

The design of the two \biceparray\ receivers at 95 and 150\,GHz is at an advanced level of development and is based on the same TDM architecture and TES detector arrays. Table~\ref{tab:ba_tmux} shows the multiplexing factors and cabling arrangement for each receiver. We are currently in the process of validating the 150\,GHz module design, which allows an 18x18 pixel tile (see Figure~\ref{fig:150GHz_module}) to be read out. The large number of MCEs needed at 150\,GHz required a relatively minor mechanical redesign of the MCE boxes in order to fit within the space constraints imposed by the telescope mount. In the context of CMB-S4~\cite{2017arXiv170602464A}, our team has developed ambitious preliminary designs of TDM modules that can read up to 1,920 detectors, without requiring superconducting flex cables.

The last \biceparray\ receiver at 220/270\,GHz will eventually feature over 20,000 detectors, and will be read out using RF-multiplexing schemes, which promise higher multiplexing factors than TDM. One possible implementation is the microwave SQUID readout ($\mu$Mux~\cite{henderson2018umux}) system that has been already been operated with our current TES detectors for a year as an on-sky demonstrator\cite{Cukierman19}. In this system, the TES detectors are unchanged and are multiplexed by coupling each TES to its own superconducting microwave resonator through an RF-SQUID. Alternatively, Thermal Kinetic Inductance Detectors (TKID~\cite{Wandui2020,Minutolo2019}) offer on-wafer RF multiplexing to greatly simplify wafer hybridization, and are currently being developed and tested within the collaboration. 

\begin{table}
  \center
  \begin{tabular}{l c c c} 
    \toprule 
    Frequency			&	30/40~GHz 	&	95~GHz	&	150~GHz			\\
    \midrule 
    \# Detector Tiles			&	12			&	12		&	12				\\
    \# Detectors		&	192 + 300	&	4,056		&	7,776				\\
	\# Detectors/Tile			&	32 + 50		&	288		&	648				\\
	\# SQUID MUX11 chips/Tile 	&	6			&	32		&	64				\\
	\# MCE 				&	1			&	3		&	6				\\
	\# Columns/MCE 		&	24			&	32		&	32				\\    
	\# Rows (multiplexing factor) 			&	33			&	43		&	41				\\  
	\# SQUID series array (SSA) modules		&	4			&	12		&	24				\\   
	\# Manganin cables (300\,K - 4\,K) 			&	5			&	15		&	30				\\    
	\# NbTi cables (4\,K - 280\,mK)		&	3			&	9		&	18				\\   
    \bottomrule
    \vspace{1mm}
  \end{tabular}
  \caption{Time-domain multiplexing (TDM) scheme for the 30/40, 95, and 150~GHz receivers in \biceparray. The fourth 220/270 GHz receiver will use microwave $\mu$MUX \cite{henderson2018umux,Cukierman19} or TKIDs~\cite{Wandui2020,Minutolo2019}, which both promise higher multiplexing factors.}
  \label{tab:ba_tmux}
\end{table}

\subsection{Magnetic Shielding}
\label{magnetic_shield}

\begin{figure}[htbp]
\begin{center}
\includegraphics[width=1.0\linewidth, keepaspectratio]{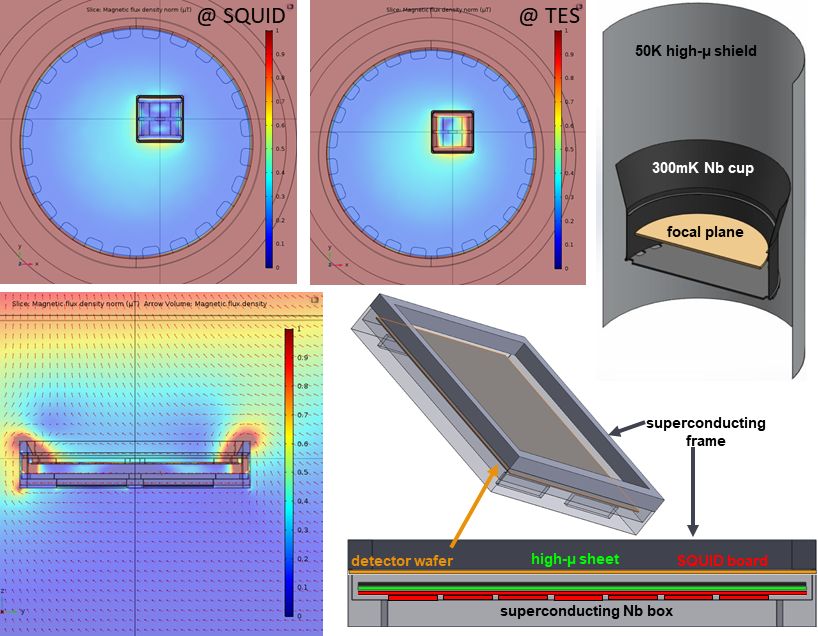}
\caption{\biceparray\ magnetic shielding. {\it Top left and center}: Comsol simulation of the residual total magnetic flux density when a purely axial $100\, \mu$T external magnetic DC field is applied. Two horizontal slices of the receiver are shown, at the SQUID and TES planes, respectively. Color bar units are in $\mu$T. {\it Bottom left}: same as above but for a vertical slice at the detector module. {\it Top Right}: A 1\,mm thick, high-$\mu$ A4K sheet is wrapped around the 50\,K stage. A superconducting niobium flared cup shield is mounted at the 300\,mK stage, surrounding the focal plane.  {\it Bottom right}: The modules provide an additional level of shielding for the SQUIDs and the detectors. The various components made out of different materials are highlighted in color for clarity, and their respective labels are colored accordingly.}

\label{ba_mag_shield}
\end{center}
\end{figure}

The SQUID/TES readout system is known to be susceptible to external magnetic fields\cite{Claycomb99,Vavagiakis18} and \biceparray's magnetic shielding architecture is designed to minimize the level of spurious signals generated as the telescope moves along the Earth's magnetic field lines. We studied multiple configurations using COMSOL Multi-Physics software~\cite{Bergen16}, which allowed us to simulate the Meissner behavior of a superconducting material (see Figure~\ref{ba_mag_shield}). We selected an architecture with a 1\,mm-thick cylindrical A4K high magnetic permeability ($\mu$) sheet wrapped around the 50\,K cryostat shell, combined with a 2\,mm-thick Niobium superconductive flared cup at 300\,mK, which provide a shielding factor of 200 or better at the focal plane. Slotted mounting holes on the 50\,K shield allow sufficient thermal expansion mismatch between the aluminum and A4K. The sub-Kelvin Niobium shield required a continuous base in addition to the flared cylinder to enhance the shielding performance. The Niobium base is not electrically connected to the copper base plate to avoid potential superconductor-metal interactions at the contact surface.

The TESs and the SQUIDs are then further shielded inside the focal plane modules, each of which features a superconducting aluminum detector frame, a superconducting Niobium SQUID box and a 0.020"-thick sheet of high-$\mu$ A4K material to provide a combined shielding factor of about 500 or better. Simulation shows the biggest field leakage comes from the slots for the readout cables. An additional Niobium sheet (not shown in Figure~\ref{ba_mag_shield} for simplicity), with cable slots that are offset with respect to those in the SQUID box, is mounted on the back side of the module Niobium housing to minimize the leakage. Table~\ref{tab:ba_mag} shows the simulated magnetic shielding performance of the receiver and module, listed separately. The combined system shielding performance is less straightforward to simulate due to interaction among the shields, and is beyond the scope of this paper.

We have performed direct measurements of the system to validate the COMSOL simulations by applying an external field that is modulated at a known frequency, including the use of cryogenic flux gate sensors installed at the location of the SQUIDs and TES. The results of these measurements will be published in a dedicated study (Schillaci, Basu Thakur et al, in prep.)

\begin{table}
  \center
  \begin{tabular}{l c c} 
    \toprule 
    Parts				&	Axial Residual Flux 	&	Transverse Residual Flux	\\
    \midrule 
    50K and 4K shields (combined)			&	0.24\%			&	0.52\%		\\
    Detector module (as a whole)				&	0.12\%			&	0.20\%		\\    
    \bottomrule
    \vspace{1mm}
  \end{tabular}
  \caption{Simulated magnetic shielding performance in \biceparray, listed as separate sub-system contributions.}
  \label{tab:ba_mag}
\end{table}

\subsection{RF Shielding}
\label{rf_shield}

Following the successful electromagnetic interference (EMI) mitigation strategy of previous small-aperture telescopes in the \bicep/\keck\ series~\cite{ade2014bicep2}, the detectors and SQUID readout chain are enclosed in a radio frequency (RF) shield. The RF shield begins on the top of the focal plane plate and extends down to the 4\,K insert volume. Circular clamps at each temperature stage hold an aluminized Mylar shroud that drapes over the ``wedding-cake'' truss structure (a photo showing this has been published in a previous conference proceeding~\cite{Schillaci19}). The continuous Mylar sheet is used to create a conductive path that surrounds the stages at different temperatures without thermally linking them (see Table~\ref{tab:ba_subK_load} for an estimate of the thermal loading due to the Mylar). The ring on the stage at the bottom of the truss connects to the aluminum walls of the fridge volume and the 4\,K base plate, which then connects to the optics tube. Filter connectors (micro-D saver PI filtered, model no. C48-00063-01~\footnote{\url{https://portal.cristek.com/admin/client-data/cristek-web-links/cwl_links/4/FILTRevD.PDF}}) at the 4\,K base plate protect the cold electronics from RF interference picked up in wiring outside the cryostat.

\section{BICEP Array 30/40\,GHz Receiver Performance}

The 30/40\,GHz BA1 cryostat was successfully integrated with the new \biceparray\ mount and cooled to base temperatures in January 2020. Table~\ref{tab:2020_temps} reports the median system temperatures in BA1 during CMB observations in 2020, demonstrating adequate cryogenic performance achieved already during the first season. The exact focal plane configuration that was deployed is described in Section~\ref{sec:detectors} and will remain unchanged for two full observing seasons due to the COVID-19 pandemic.

\begin{figure}
  \centering
  \includegraphics[width=.5\textwidth]{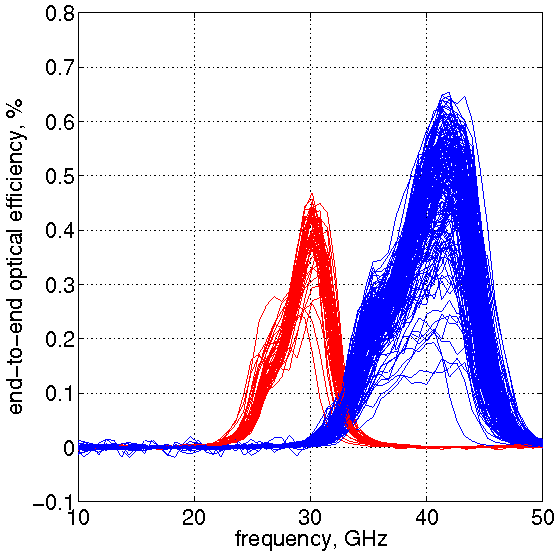}
  \caption{FTS spectra, normalized to the per-detector end-to-end optical efficiency.}
  \label{fig:FTS_spectra}
  \index{figure}
\end{figure}

An extensive detector calibration campaign was conducted over the first several months of 2020, by a combination of the Austral Summer crew and winter-over Nathan Precup. Figure~\ref{fig:FTS_spectra} shows FTS spectra for all the single-band detectors in BA1, demonstrating well-behaved band-passes at both frequencies and high end-to-end optical efficiency from the combination of detectors and optical train.

\begin{figure}
  \centering
  {\includegraphics[width=.49\columnwidth, clip=true]{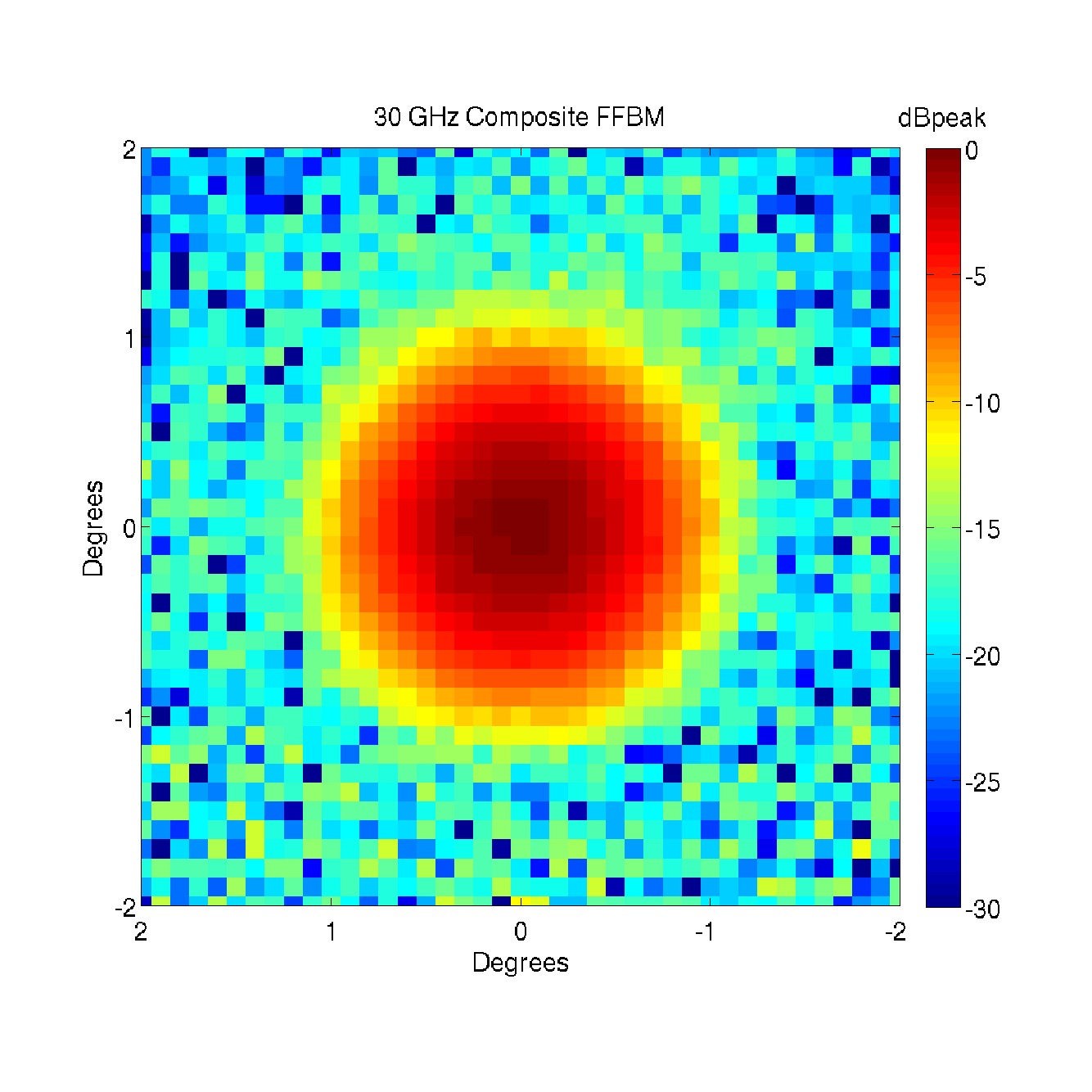}}
  {\includegraphics[width=.49\columnwidth, clip=true]{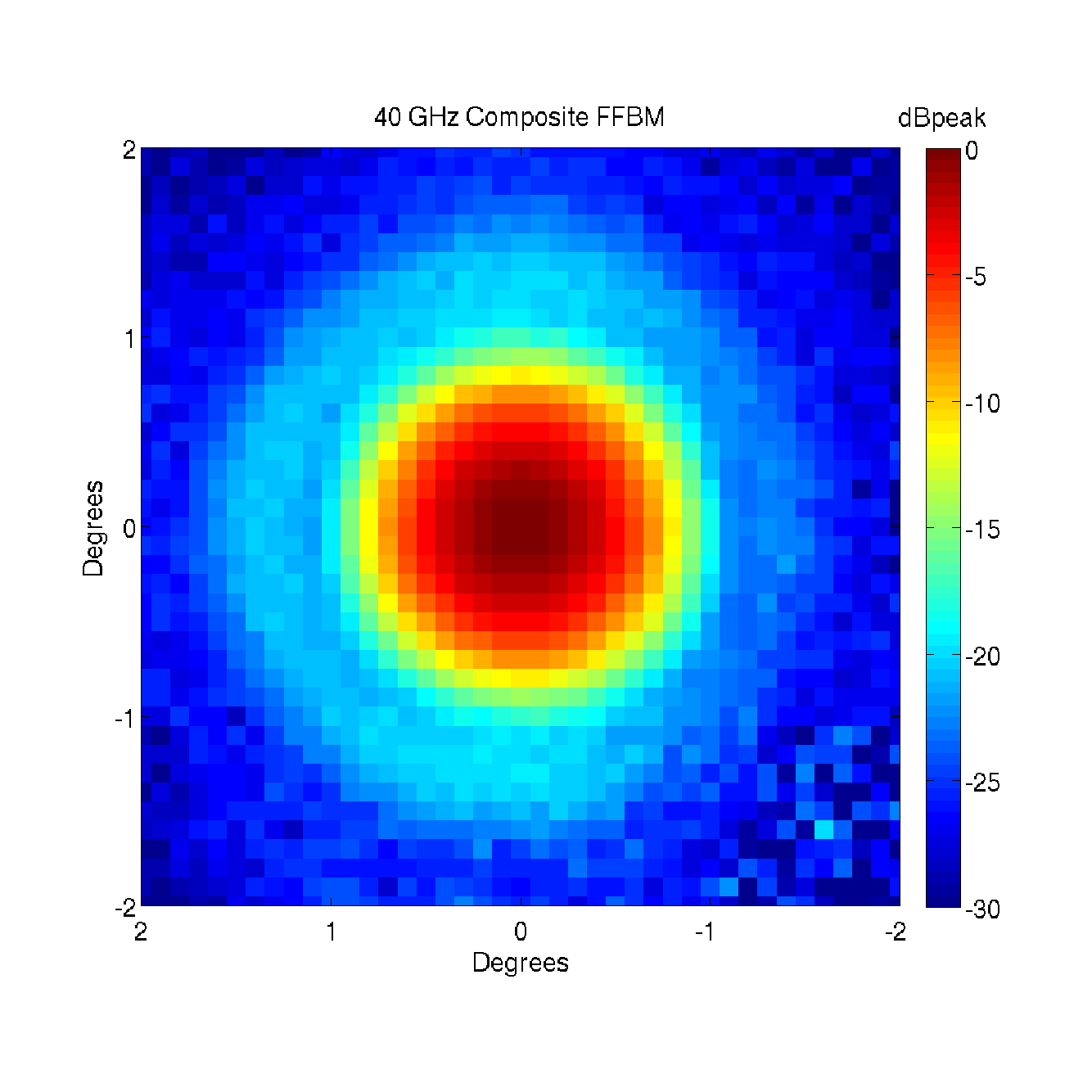}}
  \caption{Band-averaged far-field beam map (FFBM) composites, made by coadding all per-detector FFBM composites in a frequency band. Per-detector FFBM composites are made by coadding all individual FFBM for a detector, which are taken at a range of boresight angles to get maximum coverage of the source. {\it Left}: 30\,GHz. {\it Right}: 40\,GHz.}
  \label{fig:beam_maps}
  \index{figure}
\end{figure}

Another major check of the optical system is provided by taking beam maps with a source in the far-field of the telescope~\cite{Karkare2019,St.Germaine19}. Figure~\ref{fig:beam_maps} shows band-averaged far-field beam map (FFBM) composites, obtained by coadding all per-detector FFBM composites in a frequency band (not including the prototype dual-band detectors mentioned in Section~\ref{sec:detectors}). Per-detector FFBM composites are made by coadding all individual FFBMs for a detector, which are taken at different boresight angles. The band-averaged composites demonstrate in-focus imaging and FWHM beam sizes in line with the design values of 76 and 57\,arcmin at 30 and 40\,GHz, respectively (see Table~\ref{tab:rxs}). 

\begin{figure}
  \centering
  {\includegraphics[width=.49\columnwidth, clip=true]{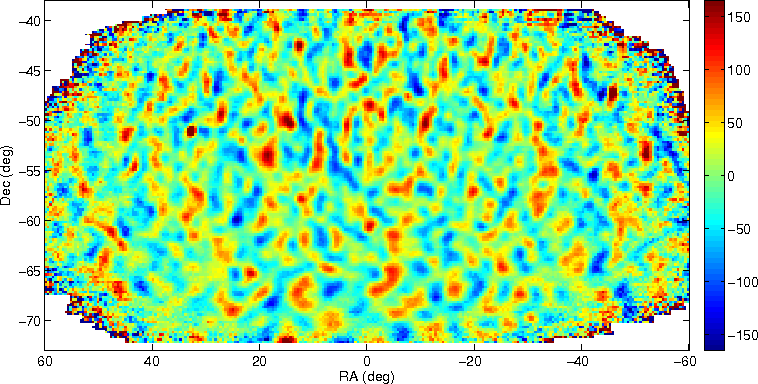}}
  {\includegraphics[width=.49\columnwidth, clip=true]{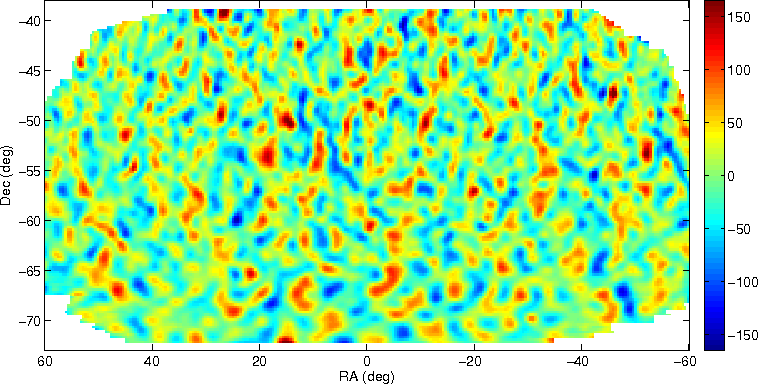}}
  \caption{{\it Left}: \biceparray\ 40\,GHz temperature map covering 570\,deg$^2$, obtained from 3,187 50-minute scansets from the 2020 observing season. {\it Right}: \planck\ 44\,GHz\cite{Planck2015_VI} reobserved temperature map, for comparison. Color bar units: $\mu$K.}
  \label{fig:40GHz_map}
  \index{figure}
\end{figure}

Finally, Figure~\ref{fig:40GHz_map} shows a preliminary \biceparray\ 40\,GHz temperature map covering 570\,deg$^2$, obtained from 3,187 50-minute scansets from the 2020 observing season, and compared to the \planck\ 44\,GHz\cite{Planck2015_VI} reobserved temperature map. Besides the variable~\cite{Fuhrmann2014} QSO [HB89] 0208-512~\footnote{\url{https://ned.ipac.caltech.edu/byname?objname=[HB89]\%200208-512}} visible at [RA, Dec] = [32.6925, -51.0172]\,deg in the \biceparray\ field, the maps look very similar. Further analysis is beyond the scope of this paper.

\begin{table}
  \center
  \begin{tabular}{l c l c} 
    \toprule 
    Thermometer				&	Temperature (K) & Thermometer				&	Temperature (K)	\\
    \midrule 
    50\,K pulse tube head & 41.803 & 4\,K heat strap cold side & 3.562\\
    50\,K heat strap cold side & 43.664& 4\,K heat strap warm side & 3.689\\
    50\,K heat strap warm side & 49.036& Fridge bracket & 3.943\\
    50\,K plate     & 50.251& Field lens & 4.434\\
    50\,K tube bottom & 58.737& Nylon filter & 4.896\\
    50\,K tube top & 63.268 & Objective lens &5.183\\
    50\,K alumina filter & 63.624 & SQUID Series Array (SSA) board &4.002\\
    \midrule
    He-4 switch & 22.154 & 2\,K plate & 2.457\\
    He-3 switch & 23.191 & Niobium cup (flared section) & 0.373\\
    He-4 pump & 5.006   & IC heat strap & 0.322\\
    He-3 pump & 4.895  & IC plate & 0.321\\
    He-4 condenser & 3.781 & Edge filter & 0.321\\
    He-4 evaporator & 2.492 & UC plate side & 0.303\\
    Intercooler (IC) evaporator & 0.318 & UC plate center & 0.302\\
    Ultracooler (UC) evaporator & 0.274 & UC heat strap & 0.279\\
    \bottomrule
    \vspace{1mm}
  \end{tabular}
  \caption{Median \biceparray\ system temperatures during CMB observations in 2020. See Figure~\ref{fig:ba_raytrace} for locations.
  }
  \label{tab:2020_temps}
\end{table}

\section{Observing Strategy and Projected Sensitivity}

\begin{figure}
  \centering
  \includegraphics[width=0.775\textwidth]{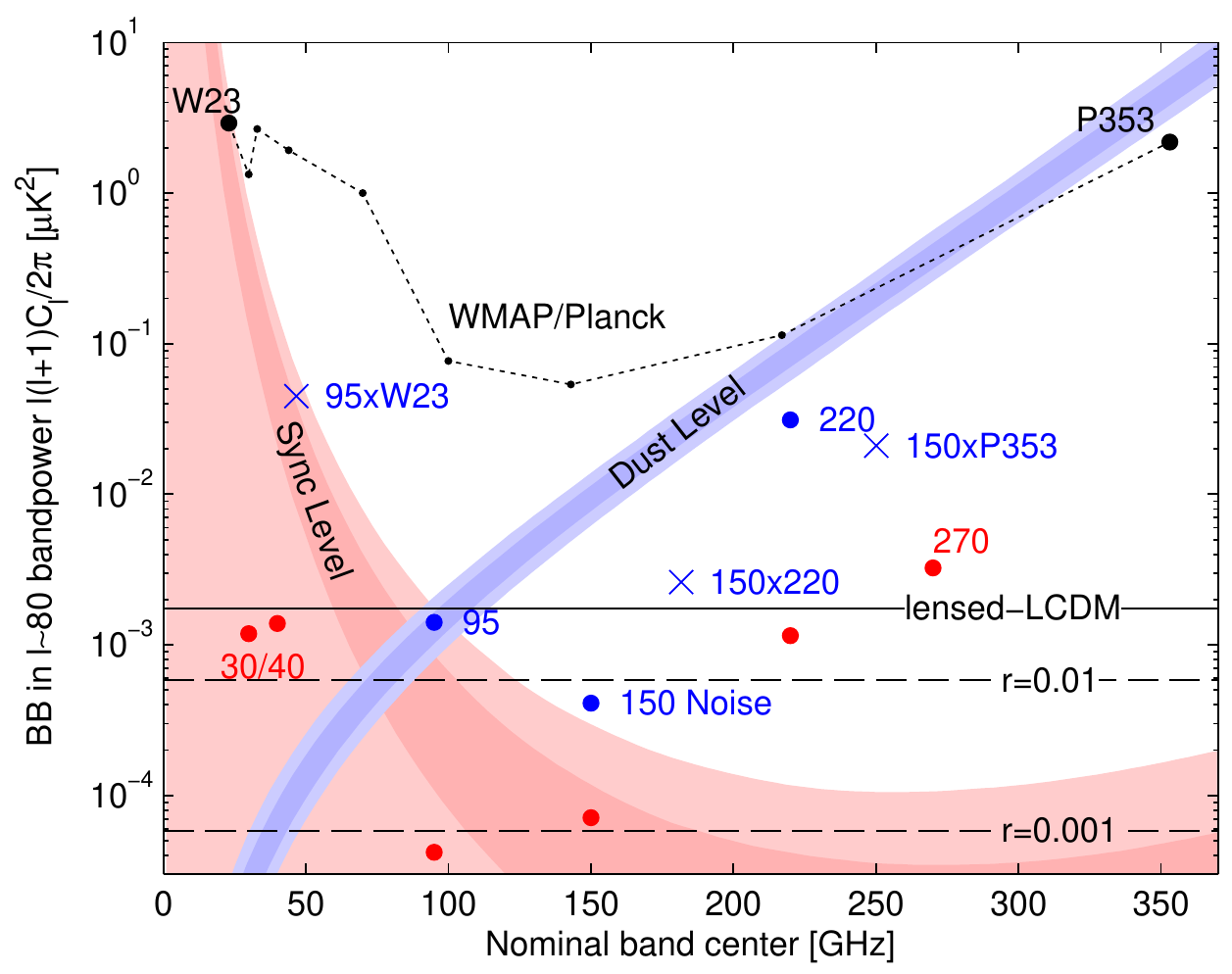}
  \caption{Noise uncertainty compared to signal level in the crucial $\ell = 80$ $BB$ bandpower in the \bicep/\keck\ sky patch. The horizontal black lines show the expected signal power of lensed-$\Lambda$CDM, $r = 0.01$ and $r=0.001$. The black circles show the noise uncertainty in the BK field of \wmap\ (23 \& 33\,GHz) and \planck\ (seven bands spanning 30--353\,GHz) auto-spectra. The blue circles and crosses show the noise uncertainties published in BK15\cite{BK15} for auto- and cross-spectra, respectively. The constraints on the dust and synchrotron signal levels set in BK15 are shown by the shaded regions. Dust emission is detected with high signal-to-noise, while synchrotron has not been detected to date. The projected \biceparray\ sensitivities by the end of the 2024 season are shown in red (see Figure~\ref{fig:bk_projections_2018} for a timeline).}
  \label{fig:noilev_bk24}
  \index{figure}
\end{figure}

\biceparray\ focuses its CMB observations on the same sky patch as \bicepthree, spanning right ascension $-60\deg<$RA$<60\deg$ and declination $-70\deg<\delta<-40\deg$, with an effective area of $\sim 600\text{deg}^2$ (larger than \keck's $\sim 400\text{deg}^2$, but colocated). 
With the full \biceparray\ sensitivity (see Figure~\ref{fig:noilev_bk24}), we expect to achieve and surpass the \planck\ map depths at all frequencies after only a few months of observations.

\begin{figure}
  \centering
  \includegraphics[width=1.\textwidth]{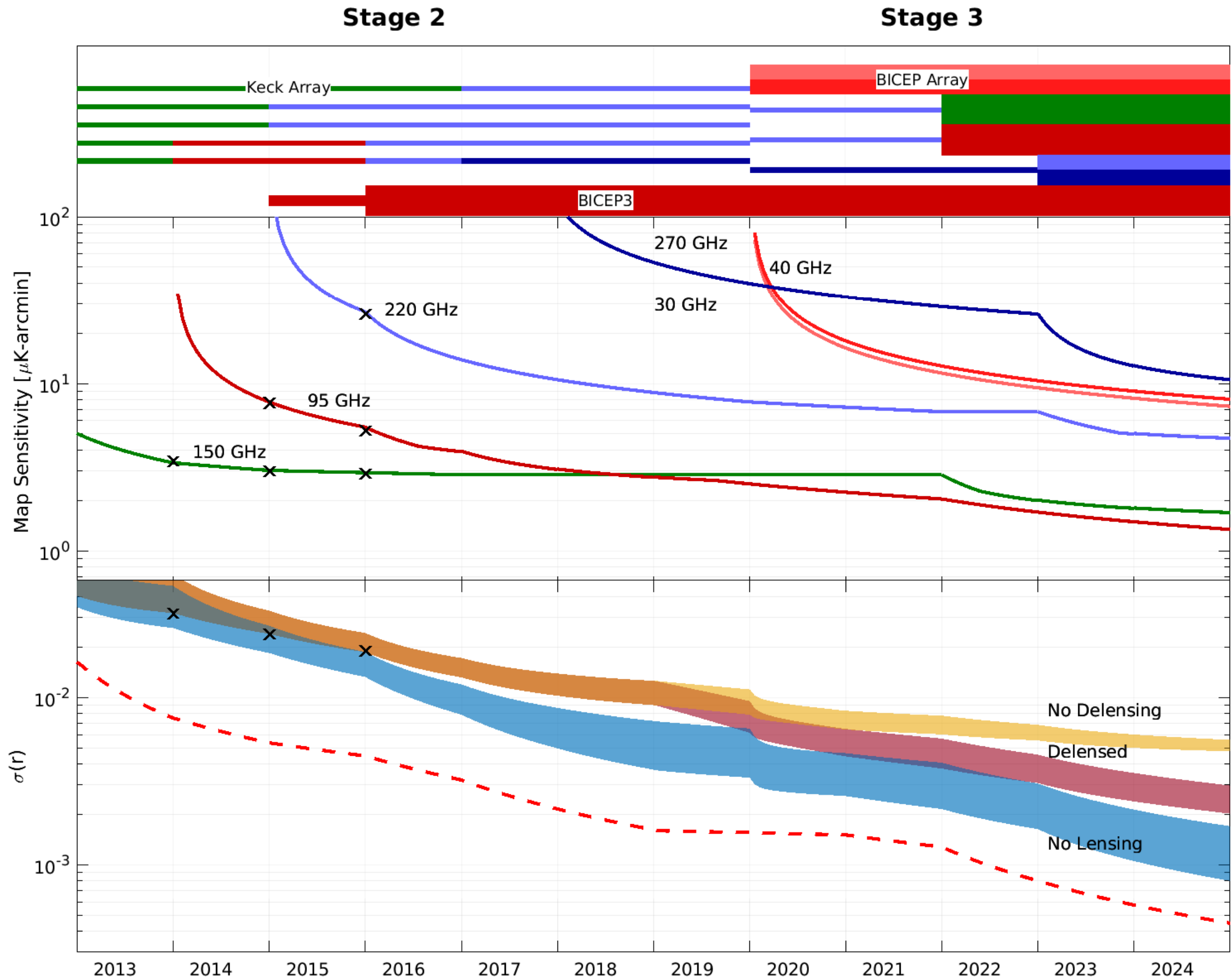}
  \caption{Projected sensitivity of the ongoing and planned \bicep/\keck\ observational program. \biceparray\ will
provide the sensitivity necessary to sustain the current rapid progress in constraining inflation. Note that these projections
involve direct scalings from published end-to-end analyses and hence include all real-world inefficiencies, and removal of dust and
synchrotron foregrounds. The X’s mark the sensitivities achieved in the BK13\cite{BicepKECKxPlanck}, BK14\cite{Keck2016} and BK15\cite{BK15} papers. Top panel: Schematic representation of the program showing the frequency bands covered by the various receivers through each observing
season. Middle panel: Map depth at each frequency as a function of time. Bottom panel: Sensitivity to $r$ after marginalizing over seven foreground parameters, as well as the ``no foreground'' raw sensitivity (red dash). The shaded regions show different levels of assumed delensing efficiency (including a ``no lensing'' case) that we can achieve by working with SPT-3G~\cite{Bender2020,Wu2020} under the umbrella of the newly-founded South Pole Observatory~\cite{SPO2020}. The width of the shaded regions is due to a combination of varying the foreground amplitudes 1\,$\sigma$ low/high from the BK15 maximum likelihood model, and turning frequency decorrelation (i.e., spatially varying spectral indices) in the foreground modeling on/off.
  }
  \label{fig:bk_projections_2018}
  \index{figure}
\end{figure}

The parameters of the \keck, \bicepthree, and \biceparray\ are given in Table~\ref{tab:rxs}. The \biceparray\ sensitivity estimates are based on achieved survey weight per year of the \keck. This procedure accounts for all real-world observing imperfections such as detector yield, cryogenic efficiency, data cuts, ground subtraction, temporal filtering, and other unexpected events that decrease the final sensitivity compared to the ideal case. Figure~\ref{fig:bk_projections_2018} shows the projected sensitivity of the ongoing and planned \bicep\ program. We expected to reach $ \sigma(r) \sim 0.003$ at the end of the program, depending on the level of delensing that can be achieved using higher-resolution data from the South Pole Telescope. A demonstration of the delensing technique was just published~\cite{Wu2020}.

\acknowledgments 

\bicep/\keck\ project (including \bicep2\, \bicep3\ and \biceparray) have been made possible through a series of grants from the National Science Foundation (including 0742818, 0742592, 1044978, 1110087, 1145172, 1145143, 1145248, 1639040, 1638957, 1638978, 1638970, 1726917, 1313010, 1313062, 1313158, 1313287, 0960243, 1836010, 1056465, 1255358) and by the Keck Foundation. The development of antenna-coupled detector technology was supported by the JPL Research and Technology Development Fund and NASA Grants 06-ARPA206-0040, 10-SAT10-0017, 12-SAT12-0031, 14-SAT14-0009, 16-SAT16-0002, \& 18-SAT18-0017. The development and testing of focal planes were supported by the Gordon and Betty Moore Foundation at Caltech. Readout electronics were supported by a Canada Foundation for Innovation grant to UBC. The computations in this paper were run on the Odyssey cluster supported by the FAS Science Division Research Computing Group at Harvard University. The analysis effort at Stanford and SLAC was partially supported by the Department of Energy, Contract DE-AC02-76SF00515. We thank the staff of the U.S. Antarctic Program and in particular the South Pole Station without whose help this research would not have been possible. Tireless administrative support was provided by Kathy Deniston, Sheri Stoll, Irene Coyle, Donna Hernandez, and Dana Volponi.

\bigskip
\bigskip
\bigskip

{\small \bibliography{main}} 
\bibliographystyle{spiebibmod} 

\end{document}